\documentclass[preprint,11pt]{article}

\usepackage{amssymb, amsmath, amsthm,graphicx, wrapfig}

\begin{document}




\title{Simulations of a model \\for the Northern Spotted Owl}


\author{Brita Jung}
\date{}
\maketitle
\begin{center}
Department of Natural Sciences/Mathematics, \AA bo Akademi University,\\ F\"anriksgatan 3 B, FIN-20500 \AA bo, Finland
\end{center}
\begin{abstract}
In this paper, a branching process model of the Northern Spotted Owl is simulated. We focus on the time until extinction. It is shown how an approximation of the model with a multivariate autoregressive process works well near the equilibrium, but does not give a good estimate of the time until extinction.  We also show that introduction of randomness in some of the parameters previously assumed to be constants shortens the time until extinction considerably.

\end{abstract}






\newtheorem{thm}{Theorem}[section]
\newtheorem{claim}[thm]{Claim}
\newtheorem{cor}[thm]{Corollary}
\newtheorem{lemma}[thm]{Lemma}
\newtheorem*{thmnonumber}{Theorem}

\theoremstyle{definition}
\newtheorem*{ex}{Example}

 
\section{Introduction}

The  Northern Spotted Owl (\emph{Strix occidentalis caurina}) inhabits parts of Oregon, Washington and California in the US, as well as parts of British Columbia in Canada. It is a medium-sized owl, described by Thomas et al. in \cite{Tho} to have a "dark brown coloring with whitish spots on the head and neck, and white mottling on the abdomen and breast". The species inhabits old-growth forests, and is thus affected by logging. Efforts to manage the population began in 1977. As Thomas et al. say in \cite{Tho}, "the spotted owl issue is, to some degree, a surrogate for the old-growth issue". That is, protecting the owl is really about protecting the forests. 

We will study a bivariate branching process that models the owl population. We are mainly interested in the event that the population dies out. In section 4, computer simulation is used to compare the time until extinction for the original process with the extinction time for some approximations of it. 
\pagebreak
\begin{wrapfigure}{r}{0.45\textwidth}
  \vspace{-0pt}
  \centering
    \includegraphics[width=0.40\textwidth]{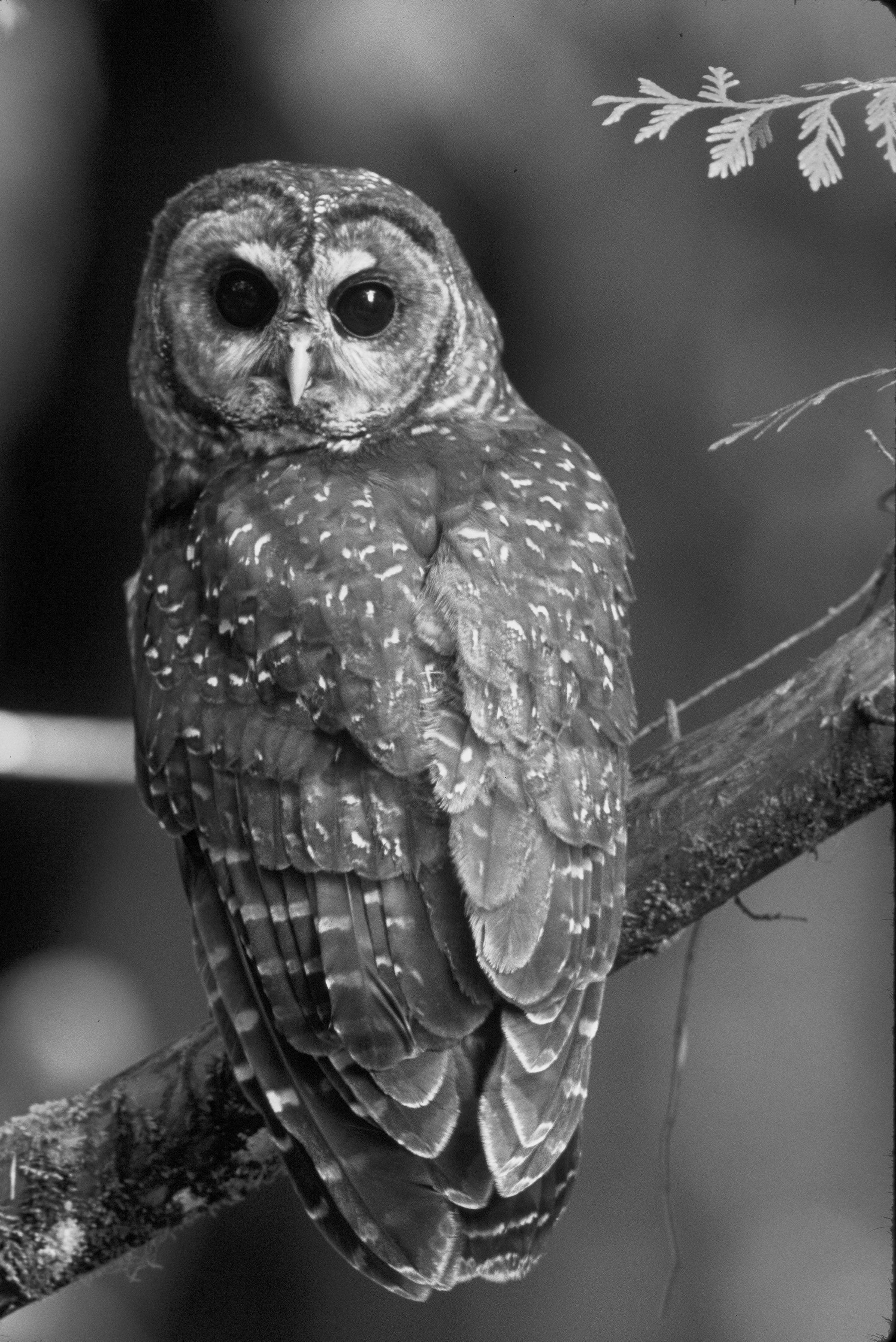}
  \caption{The Northern Spotted Owl \footnotesize (Photograph by U.S. Fish and Wildlife Service, J. and K. Hollingsworth.)}
  \vspace{-30pt}
\end{wrapfigure}
The model includes an Allee effect, that is, at low population sizes, the population growth declines. We will see that this is essential when approximating the time until extinction.

In section 5, we introduce stochasticity in some of the parameters of the branching process model, to see what effect this has on the extinction time.

In their report, Thomas et al. (\cite{Tho}) mention several factors that one must take into account when formulating a conservation stategy. Some of these are laws and regulations, land ownership, and regional and national cultures. In this paper we will not deal with any of these factors, but focus only on the mathematical model.


\section{A deterministic model for the Northern Spotted Owl}

The following deterministic discrete time model of the owl population was first introduced by Thomas et al. in \cite{Tho}. We use the formulation and notation of Allen et al. (\cite{All}). The model consists of five state variables and a number of auxiliary variables and parameters, four state equations and five auxiliary equations. The state variables, the auxiliary variables and the parameters are given in tables 1-3 below, where $t$ denotes the time in years.

\begin{table}[!ht]
\centering
\begin{tabular}{|ll|}
\hline 
 & \\
$J_t$ & Number of juveniles at time $t$\\
$S_t$ & Total number of single adults at time $t$\\
$P_t$ & Number of pairs at time $t$\\
$S_{m,t}$ & Number of single males at time $t$\\
$S_{f,t}$ & Number of single females at time $t$\\ 
& \\
\hline
\end{tabular}
\caption{The state variables}
\label{tab:statevariables}
\end{table}

\begin{table}[!ht]
\centering
\begin{tabular}{|ll|}
\hline
 & \\
$O_t$ & Number of occupied sites at time $t$\\
$A_t$ & Number of available sites at time $t$\\
$U_t$ & Number of suitable sites at time $t$, $U_t\le T$\\
$D_t$ & Probability of juveniles surviving dispersal at time $t$\\
$M_t$ & Probability of a female finding a male at time $t$\\
 & \\
\hline
\end{tabular}
\caption{Auxiliary variables}
\label{tab:auxiliaryvariables}
\end{table}

\begin{table}[!ht]
\centering
\begin{tabular}{|lll|}
\hline
& &\\
$s_S$ & Fraction of single owls surviving one year & 0.71\\
$s_J$ & Fraction of juveniles surviving to single adults in one year & 0.60\\
$p_s$ & Probability that a pair survives one year and does not split & 0.88\\
$p_b$ & Probability that a pair survives one year and splits & 0.056\\
$f$ & Number of offspring per breeding pair in one year & 0.66\\
$m$ & Unoccupied site search efficiency & 20\\
$n$ & Unmated male search efficiency & 20\\
$T$ & Total number of sites & 1000\\
& &\\
\hline
\end{tabular}
\caption{The parameter values used.}
\label{tab:parametervalues}
\end{table}

\bigskip\bigskip

The state equations are
\begin{align}
J_t &= P_tf\\
P_t &= P_{t-1}p_s + S_{m,t-1}s_SM_{t-1}\\
S_{m,t} &= \frac12 J_{t-1}s_JD_{t-1} + S_{m,t-1}s_S(1-M_{t-1})+p_bP_{t-1}\\
S_{f,t} &= S_{m,t},
\end{align}
where $t = 0,1,2,\ldots $ denotes the time in years. The following auxiliary equations are used:
\begin{align}
S_t &= S_{m,t} + S_{f,t} = 2S_{m,t}\\
O_t &= P_t + S_{m,t}
\end{align}
\begin{align}
A_t &= \max\{0,U_t-O_t\}\\
D_t &= 1-(1-\frac{A_t}{T})^m\\
M_t &= 1-(1-\frac{\min \{S_t,T \}}{T})^n.
\end{align}
As Allen et al. (\cite{All}) say, the assumptions on the state variables and the auxiliary variables show that the system can actually be described with only two state variables, the number of pairs $P_t$ and the number of single males $S_{m,t}$. We then have the dynamical system
\begin{align}
\label{Fdef} P_{t+1} &= P_tp_s + S_{m,t}s_SM_t =: F(P_t,S_{m,t})\\
\label{Gdef} S_{m,t+1} &=  P_t(\frac12fs_JD_t +p_b)+ S_{m,t}s_S(1-M_t) =: G(P_t,S_{m,t}),
\end{align}
where $M_t$ and $D_t$ are as above. (Note that $M_t$ and $D_t$ depend on $P_t$ and $S_{m,t}$, so the system is not at all linear.) The system has fixed points $(\bar P,\bar S_m)$ that are solutions of the equation
\begin{equation}
\left\{ \begin{array}{clc}
\bar P &=& F(\bar P,\bar S_m)\\
\bar S_m &=& G(\bar P,\bar S_m),
\end{array}\right.
\end{equation}
where $F$ and $G$ are as in equalities \ref{Fdef} and \ref{Gdef}. Since $F$ and $G$ are complicated functions, the equilibria cannot be calculated explicitly, but we can get numerical solutions. The number of suitable sites $U_t$ will be kept constant, $U_t = U$. Allen et al. (\cite{All}) have shown, that for $U<149$ the only equilibrium is the origin, which means certain extinction. At $U\approx 149$, there is a bifurcation and when $U> 149$, there are three equilibria. 

For our study, we use $U=160$, so that the effect of logging is constant and rather large and the number of available sites rather small. In this case, the three fixed points are
\begin{equation}
E_0 = (0,0), E_1 \approx (34.0116,13.5771)\mbox{  and  }  E_2 \approx (71.4424,20.9764).
\end{equation}
The Jacobian matrix of the system is $J$, where
\begin{align*}
J_{11} &= \frac{\partial F}{\partial P} = p_s \\
J_{12} &= \frac{\partial F}{\partial S_m} =s_S\left(1+\frac{2nS}{T}\left(1-\frac{2S}{T}\right)^{n-1}-(1-\frac{2S}{T})^n\right)\\ 
J_{21} &= \frac{\partial G}{\partial P} = \frac f2 s_J\left(1\!-\!\frac{mP}{T}\left(1-\frac{U-P-S}{T}\right)^{m-1}\!\!\!\!\!\!\! - \left(1\!-\!\frac{U-P-S}{T}\right)^m\right) + p_b\\
J_{22} &= \frac{\partial G}{\partial S_m} = -\frac 12 fs_JP\left(\frac{m}{T}\left(1-\frac{U-P-S}{T}\right)^{m-1}\right)\\
&  + s_S\left(\frac{-2nS}{T}\right) \left(1-\frac{2S}{T}\right)^{n-1}
 +\left(1-\frac{2S}{T})^n\right).
\end{align*}
In the fixed points, the Jacobian matrix has the values
\begin{align*}
J_{E_0} &= \left(\begin{array}{cc} 
					0.88 & 0\\ 0.2479 & 0.71 \end{array} \right), \mbox{ which has eigenvalues } 0.88 \mbox{ and } 0.71\\
J_{E_1} &= \left(\begin{array}{cc} 
					0.88 & 0.5291\\ 0.2218 & 0.1669 \end{array} \right), \mbox{ which has eigenvalues } 1.0179 \mbox{ and } 0.0290\\
J_{E_2} &= \left(\begin{array}{cc} 
					0.88 & 0.6725\\ 0.1303 & -0.037 \end{array} \right), \mbox{ which has eigenvalues } 0.967 \mbox{ and } -0.125.\\
\end{align*}
We see that $E_0$ and $E_2$ are locally asymptotically stable (since the eigenvalues of the Jacobian are smaller than one in absolute value) and that $E_1$ is unstable (it has an eigenvalue that is larger than one). For most starting points, the system ends up in either $E_0$ or $E_2$.

\begin{figure}
  \centering
    \includegraphics[width=\textwidth]{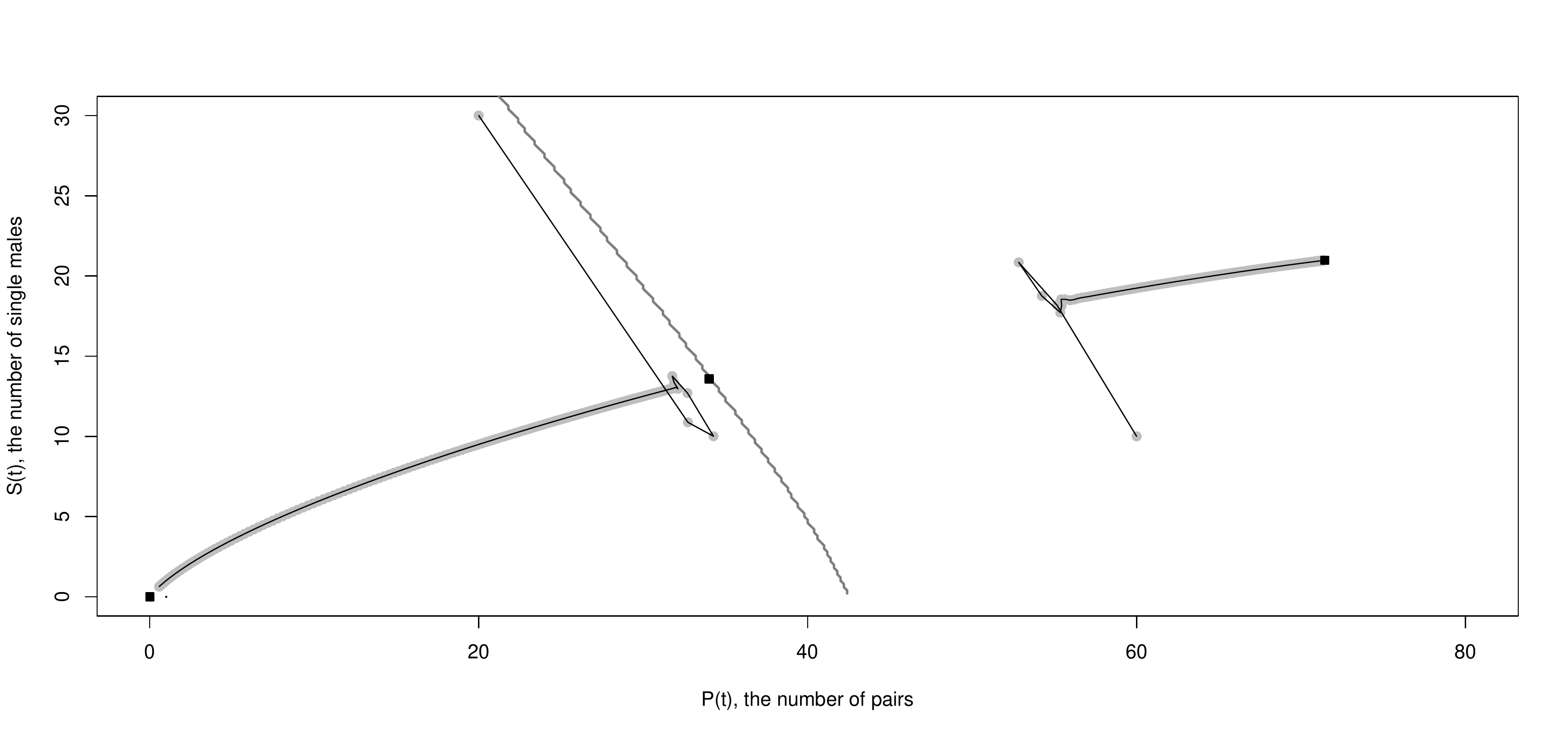}
  \caption{Two sample paths of the deterministic process. The fixed points are marked with squares, and the curve is the boundary of the stable sets of the attracting fixed points.}
\end{figure}


\section{A stochastic model for the Northern Spotted Owl}

In Allen et al. (\cite{All}) a stochastic version of the deterministic model is formulated. It is a multitype branching process. The same notation as in the deterministic case is used. $P_t$ and $S_{m,t}$ denote the numbers of pairs and single males at time $t$, respectively. The following probabilities are used:
\begin{eqnarray*}
p_s &=& \mbox{ probability that a pair survives one year}\\
& &  \mbox{ and does not split}\\
p_b &=& \mbox{ probability that a pair survives one year and splits}\\
s_SM_t &=& \mbox{ probability that a single male survives and becomes paired} \\
s_S(1-M_t) &=& \mbox{ probability that a single male survives}\\
& & \mbox{ and stays single}\\
\frac12 s_JD_tf &=& \mbox{ probability that a pair gives birth to a new male}\\
& & \mbox{ that survives and disperses to a new site.}
\end{eqnarray*}

Since $M_t$ and $D_t$ depend on $(P_t,S_{m,t})^T$, these probabilities change in every time step. The branching process that describes the population is
\begin{equation}\label{branching}
(P_{t+1},S_{m,t+1})^T = \sum_{i=1}^{P_{t}} (\eta, \theta)_i^T + \sum_{i=1}^{P_{t}} (\eta^\prime, \theta^\prime)_i^T +\sum_{i=1}^{S_{m,t}} (\eta^{\prime\prime}, \theta^{\prime\prime})_i^T,
\end{equation}
where $\{(\eta, \theta)_i^T\}_{i\ge 1}$, $\{(\eta^\prime, \theta^\prime)_i^T\}_{i\ge 1}$ and $ \{(\eta^{\prime\prime}, \theta^{\prime\prime})_i^T\}_{i\ge 1}$ are sequences of independent and identically distributed random variables distributed according to the following:
\begin{eqnarray}
(\eta,\theta)_1^T &=& \left\{\begin{array}{l} (1,0) \mbox{ with probability } p_s,\\
			 (0,1) \mbox{ with probability } p_b,\\
			 (0,0) \mbox{ with probability } 1-p_s-p_b,\end{array}\right.\\
 (\eta^\prime, \theta^\prime)_1^T &=& \left\{\begin{array}{l} (0,1) \mbox{ with probability } \frac12 s_JD_tf,\\
			 (0,0) \mbox{ with probability } 1-\frac12 s_JD_tf,\end{array}\right.\\
 (\eta^{\prime\prime}, \theta^{\prime\prime})_1^T &=& \left\{ \begin{array}{l} 
			(1,0) \mbox{ with probability } s_SM_t,\\
			 (0,1) \mbox{ with probability } s_S(1-M_t),\\
			 (0,0) \mbox{ with probability } 1-s_S. \end{array}\right.
\end{eqnarray}
For this branching process, one can show that
\begin{equation}
E((P_{t+1},S_{m,t+1})^T  | (P_t,S_{m,t})^T) = (F(P_t,S_{m,t}),G(P_t,S_{m,t}))^T,
\end{equation}
where $F$ and $G$ are the functions defined in equalities \ref{Fdef} and \ref{Gdef}. That is, the conditional expectation of the next step, given the current one, follows the deterministic model. The covariance matrix of $(P_{t+1},S_{m,t+1})^T$ given $(P_t,S_{m,t})^T$ is 
\begin{equation}\label{cov}
C =  \left( \begin{array}{cc} 
			C_{11} & C_{12}\\
			C_{21} & C_{22} \end{array}\right),
\end{equation}
where
\begin{eqnarray}
\nonumber C_{11} &=& P_tp_s(1-p_s) + S_{m,t}s_SM_t(1-s_SM_t)\\
\label{covvalues} C_{12} = C_{21} &=& -P_tp_bp_s-S_{m,t}s_S^2M_t(1-M_t)\\
\nonumber C_{22} &=& P_t\left(p_b(1-p_b)+\frac 12 s_JD_tf(1-\frac 12 s_JD_tf)\right) \\
\nonumber & & + S_{m,t}s_S(1-M_t)(1-s_S(1-M_t)).
\end{eqnarray}


\section{Simulation of the extinction time}
The deterministic model had two attracting fixed points, and for most starting points, the population ended up in one of them. The stochastic model, on the other hand, has a single absorbing state at the origin, so extinction is certain, since the process is a finite-state Markov chain and the origin is accessible from all states. This is actually quite realistic, because we do not expect any real-life population to live forever. The question is now for how long the population will live before it dies out.

The expected time until extinction cannot be calculated explicitly, so we will use computer simulation. In many cases, simulation of a branching process is slow and one needs to approximate the process by some simpler process. In this case it happens that the branching process can be easily simulated. Therefore, we will both simulate the model directly and simulate some approximations. We can then compare the results.

\subsection{Direct simulation}
\begin{figure}[!h]
  \centering
    \includegraphics[width=\textwidth]{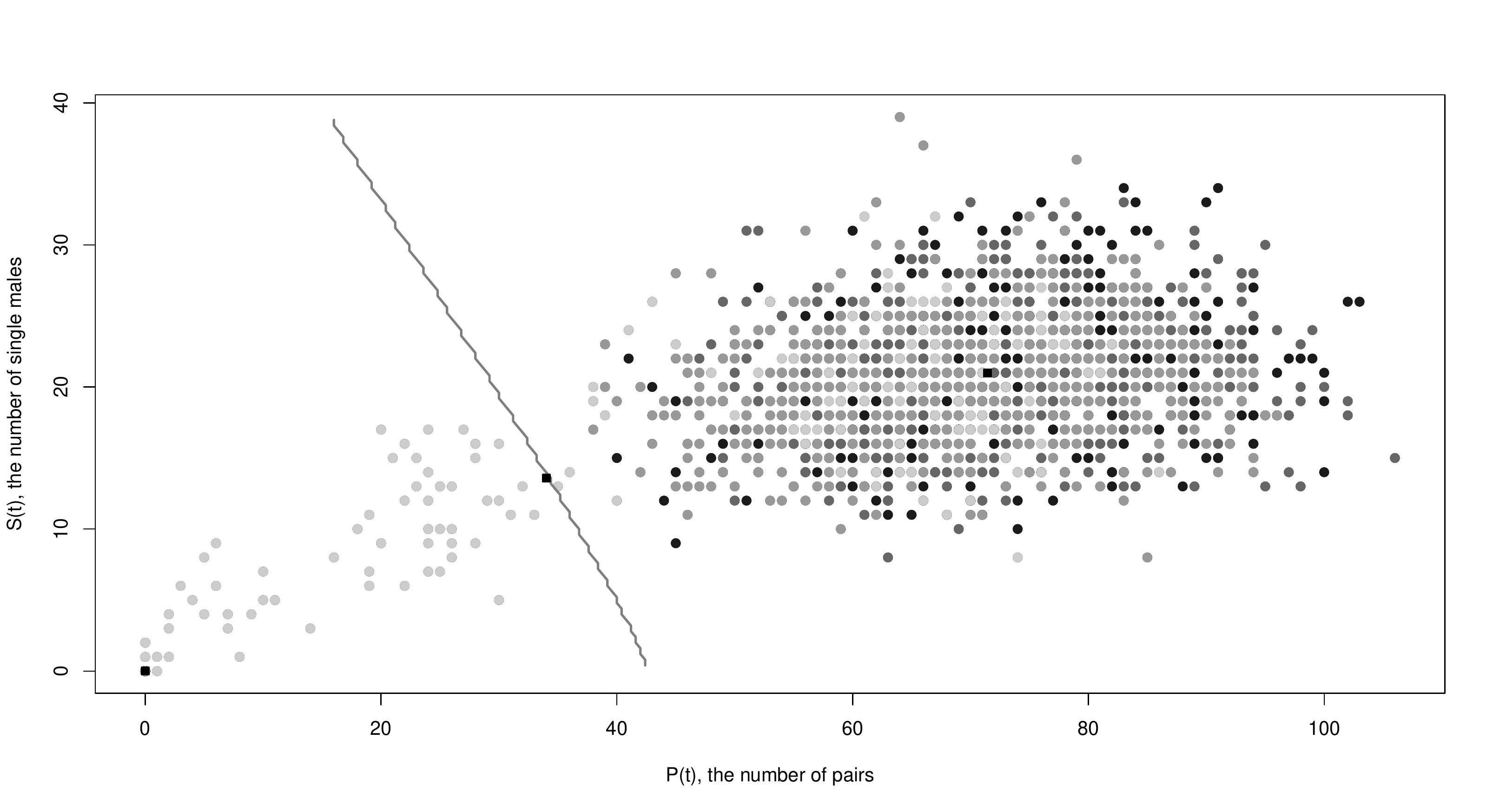}
  \caption{A sample path of the branching process, starting in $E_2$. The points get lighter gray as time passes. Extinction takes place at $t=1853$. The fixed points are marked with squares, and the curve is the boundary of the stable set of the attracting fixed points in the deterministic case.}
  \label{fig:branching}
\end{figure}

Since the distributions of the vectors $(\eta, \theta)_1^T$, $(\eta^\prime, \theta^\prime)_1^T$ and $ (\eta^{\prime\prime}, \theta^{\prime\prime})_1^T$ are so simple, we can simulate the process just by drawing a few samples from the binomial distribution in every time step. Consider the sums in equation \ref{branching}. At the time $t$, the number of paired males that survive is $T_1 := \mathrm{Bin } (P_t,p_s+p_b)$. Of these, the number of pairs that remain is $T_2 = \mathrm{Bin} (T_1 , \frac{p_s}{p_s+p_b})$ and the number of pairs that give rise to a single male is $(T_1-T_2)$. The number of single males that survive is $T_3 = \mathrm{Bin}(S_{m,t}, s_S)$ and of these, the number of new pairs formed is $T_4 = \mathrm{Bin}(T_3,M_t)$ and the number of singles that remain single is $T_3-T_4$. The number of new single males is $T_5 = \mathrm{Bin}(P_t,\frac12 s_JD_tf)$. A sample path simulated in this way can be seen in figure \ref{fig:branching}.

\subsection{Normal approximation in every step}
\begin{figure}[!h]
  \centering
    \includegraphics[width=\textwidth]{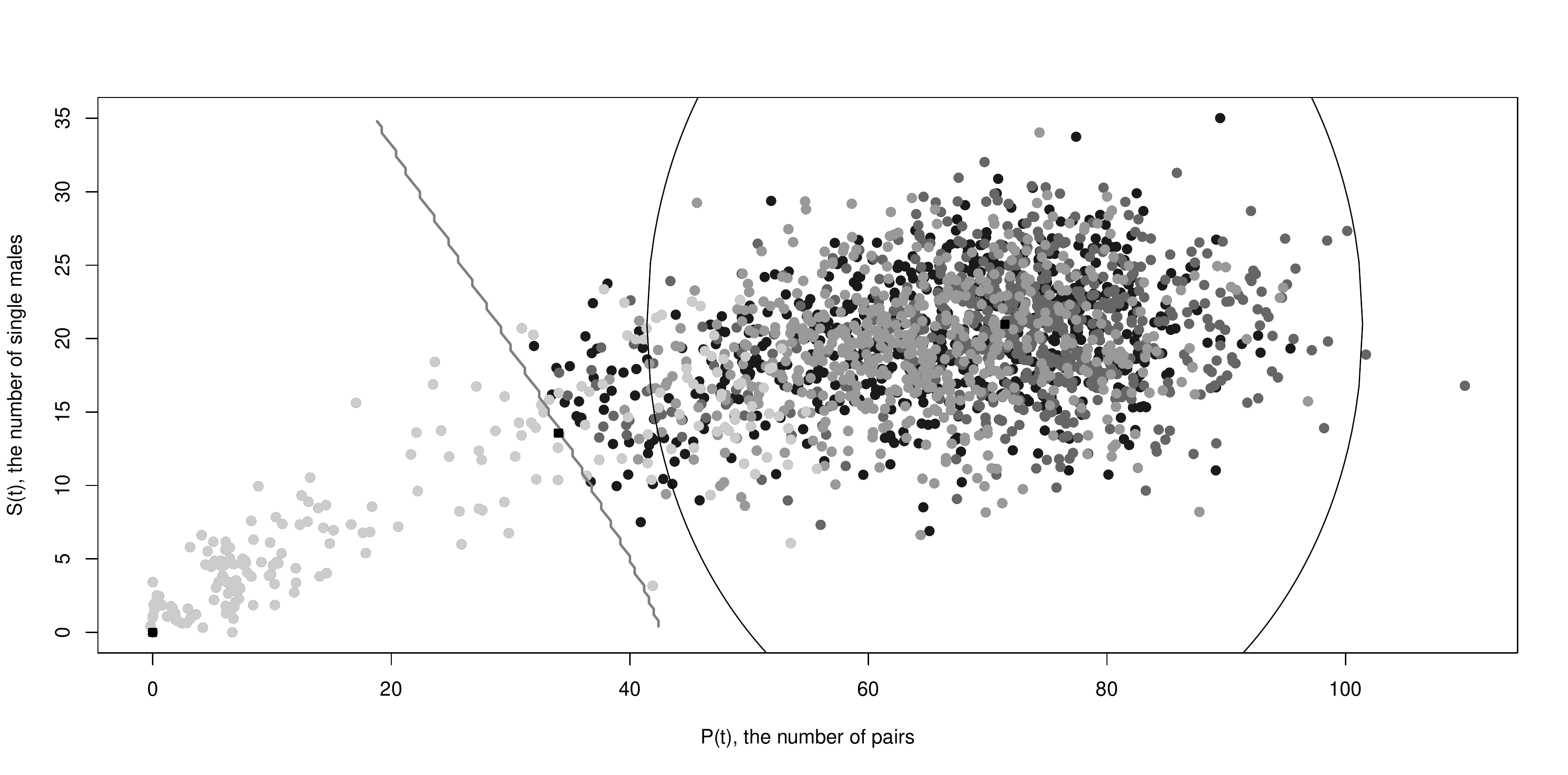}
  \caption{A sample path of normal approximation in every step, starting in $E_2$, with the first exit from a circle around $E_2$ with radius 30 at $t=212$, and extinction at $t=2021$.}
  \label{fig:normal}
\end{figure}

By the central limit theorem, a sum of independent and identically distributed random vectors, with finite covariance matrix, converge in distribution to a multivariate normal distribution. Because of this, it is reasonable to approximate a sum by the normal distribution that has the same mean and covariance matrix. This is a fairly good approximation when the number of terms in the sum is rather large. Obviously, it is not as good, when the number of terms is small. However, the right mean and covariance matrix are still used. We therefore approximate the branching process by using the normal distribution in every time step. That is, given $(P_t,S_{m,t})^T$, we let
\begin{equation}
(P_{t+1},S_{m,t+1})^T \sim N\left( (F(P_t,S_{m,t}),G(P_t,S_{m,t}))^T, C\right),
\end{equation}
where $C$ is the covariance matrix given in equations \ref{cov} and \ref{covvalues}. This process then has the same conditional expectation and covariance as the branching process. A sample path, using this approximation, is shown in figure \ref{fig:normal}.

\subsection{Approximation with an autoregressive process}
Since there are some theoretical results for exit times for multivariate autoregressive processes (see \cite{Jun}), it is of interest to try this approximation as well.

\noindent
A bivariate autoregressive process can be defined as
\begin{equation}
X_{t+1} = AX_t + \Sigma\xi_{t+1},
\end{equation}
where $A$ is a $2\times 2$ matrix and $\{\xi_t\}_{t\ge 1}$ is a sequence of independent and identically distributed bivariate normal random variables with mean $(0,0)^T$ and covariance matrix $I$ (the identity matrix).

\begin{figure}[!ht]
  \centering
    \includegraphics[width=\textwidth]{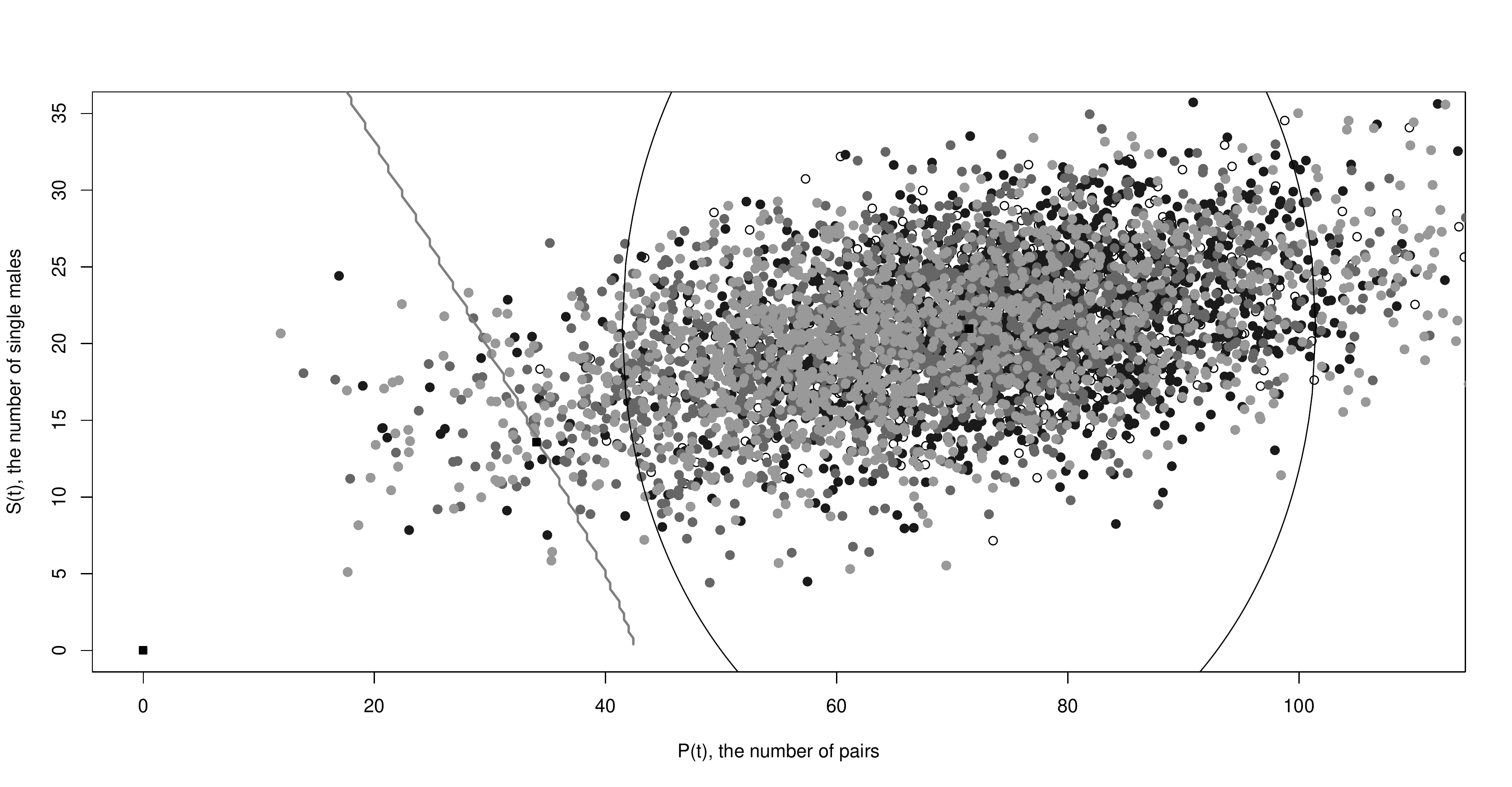}
  \caption{The autoregressive process, starting in $E_2$. A sample path of 5000 generations with the first exit from the circle at $t=444$. No extinction so far.}
  \label{arpath}
  \vspace{30pt}
\end{figure}

We want a process that is centered around the stable equilibrium $E_2 = (71.4424 , 20.9764)$, so we use $E_2$ as a starting point and let
\begin{equation}
X_t = (P_t,S_{m,t})^T - E_2.
\end{equation}
We recall that for the branching process,
\begin{equation}
E((P_{t+1},S_{m,t+1})^T  | (P_t,S_{m,t})^T) = (F(P_t,S_{m,t}),G(P_t,S_{m,t}))^T.
\end{equation}
This cannot be achieved with an autoregressive process. However, near $E_2$, $(F(P_t,S_{m,t}),G(P_t,S_{m,t}))^T$ can be approximated by $J_{E_2}((P_t,S_t)^T - E_2)$ (a Taylor expansion around $E_2$), where $J_{E_2}$ is the Jacobian of $(F(P_t,S_{m,t}),G(P_t,S_{m,t}))^T$ in the point $E_2$. Thus, we choose 
\begin{equation}
A = J_{E_2} = \left(\begin{array}{cc} 
					0.88 & 0.6725\\ 0.1303 & -0.037 \end{array} \right)
\end{equation}
in the autoregressive process. The matrix $\Sigma$ is chosen so that the covariance in the first step of the process coincides with that of the branching process. Thus, $\Sigma$ is such that
\begin{equation}
\Sigma\Sigma^T = \left( \begin{array}{cc} 
			12.6118 & -6.1030\\
			-6.1030 & 17.2571 \end{array}\right).
\end{equation}
A sample path of this autoregressive process is shown in figure \ref{arpath}.

\subsection{Comparison of the branching process simulation and the approximations}

The illustrations already suggest that normal approximation in every time step follows the original process nicely, and that the autoregressive process may be too long-lived. In each of the cases, we now simulate  5000 populations, and consider both exits from a circle with radius 30 around the fixed point $E_2$, and the time until extinction of each process. The  results are given in table \ref{tab:simresults}.

\begin{table}[!ht]
\centering
\begin{tabular}{|l|c|c|}
\hline 
 & Mean exit time from circle & Mean extinction time\\
 \hline\hline
Direct simulation & 300.81 & 1647.28 \\
Normal in every step & 299.12 & 1681.76 \\
Autoregressive & 303.31 & $>$ 1 000 000\\
\hline
\end{tabular}
\caption{Simulation results}
\label{tab:simresults}
\end{table}

The Allee effect of the original process ensures that once the process gets close to dying out, it is likely to do so. The normal approximation in every time step also includes this effect, since the conditional expectation and covariance matrix are the same as the original ones in every step. The autoregressive process, on the other hand, is a good approximation near the equilibrium $E_2$, but it contains no Allee effect, and thus the resulting extinction time is much too large.

\medskip
The bivariate autoregressive process is easy to simulate, and simple enough to allow some analytical results as well (see \cite{Jun}). However, our experiments show that one must be careful when using it as an approximation of a complicated process.

\section{Randomness in parameters}
In this section, we simulate the branching process model, with the addition that one of the parameters is assumed not to be constant, but random.

\subsection{Mean number of offspring varies}
\begin{figure}[!ht]
\vspace{-10pt}
  \centering
    \includegraphics[width=\textwidth]{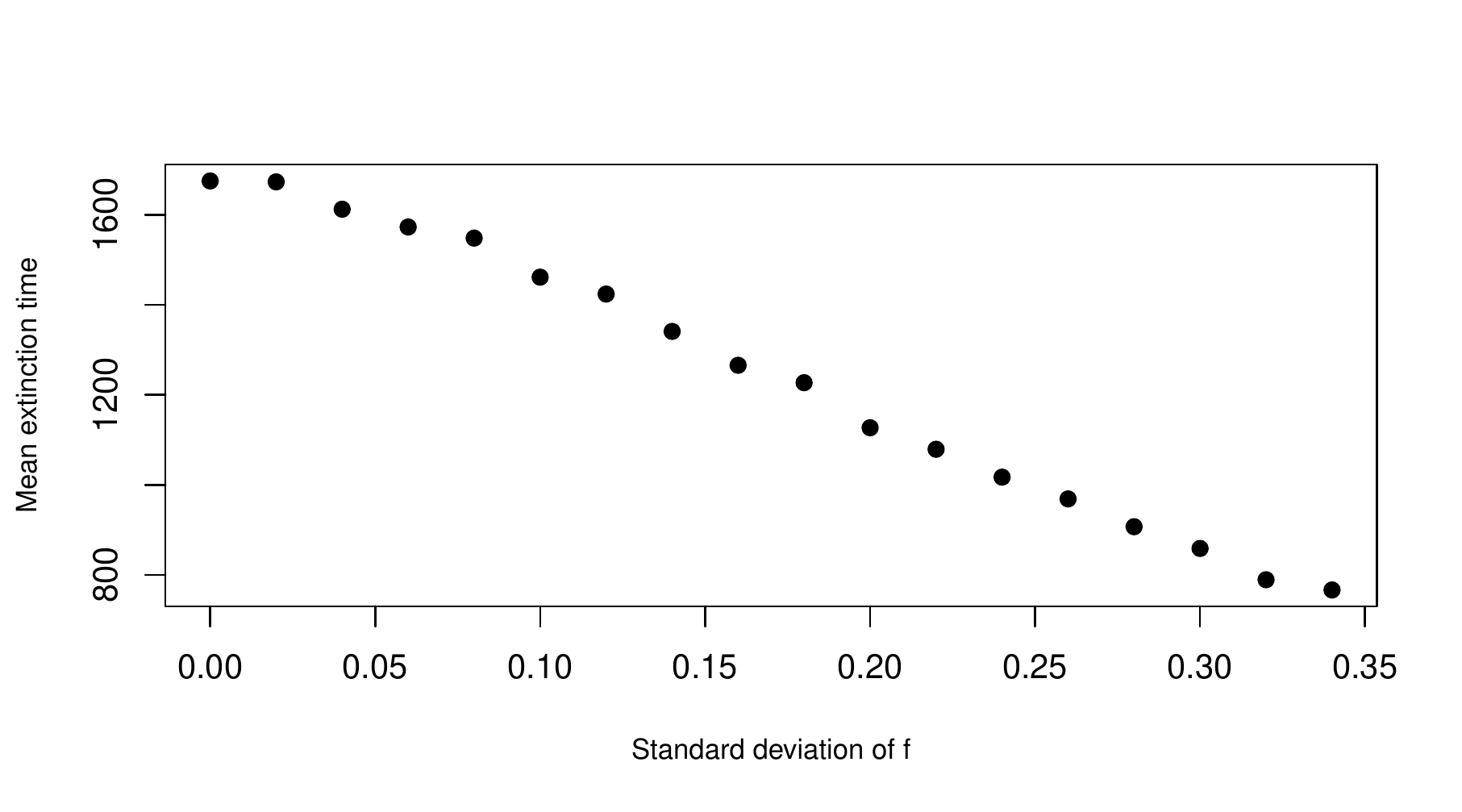}
  \caption{The mean extinction time of the branching process as a function of the standard deviation of $f$.}
  \label{Fig:f_random}
\end{figure}

In the previous sections, $f$, the mean number of offspring per breeding pair in one year, was assumed to be constant, $f=0.66$. It is likely that this number might vary over the generations. In \cite{Cou}, Courtney et al. give means and standard errors for the fecundity of the owl (their model is a different one, however). We will now try a version of the branching process model, where the mean number of offspring is not constant, but varies, in different generations. A simple random variable is used: $f$ is now assumed to be either $0.66-x$ or $0.66+x$, each with probability $1/2$, for $x = 0, 0.2, \ldots , 0.34$. Thus, for $x=0$, we have the original case, and when $x$ increases, the variance of $f$ increases. The mean of $f$ over time is still $0.66$. 5000 populations are simulated in each case, and the mean extinction time is noted. The result is plotted in figure \ref{Fig:f_random}. We see that the variation of the parameter $f$ has an impact on the extinction time. The mean extinction time seems to decrease linearly, as the standard deviation of $f$ increases. 

\subsection{Survival rate varies}
\begin{figure}[!ht]
  \centering
    \includegraphics[width=\textwidth]{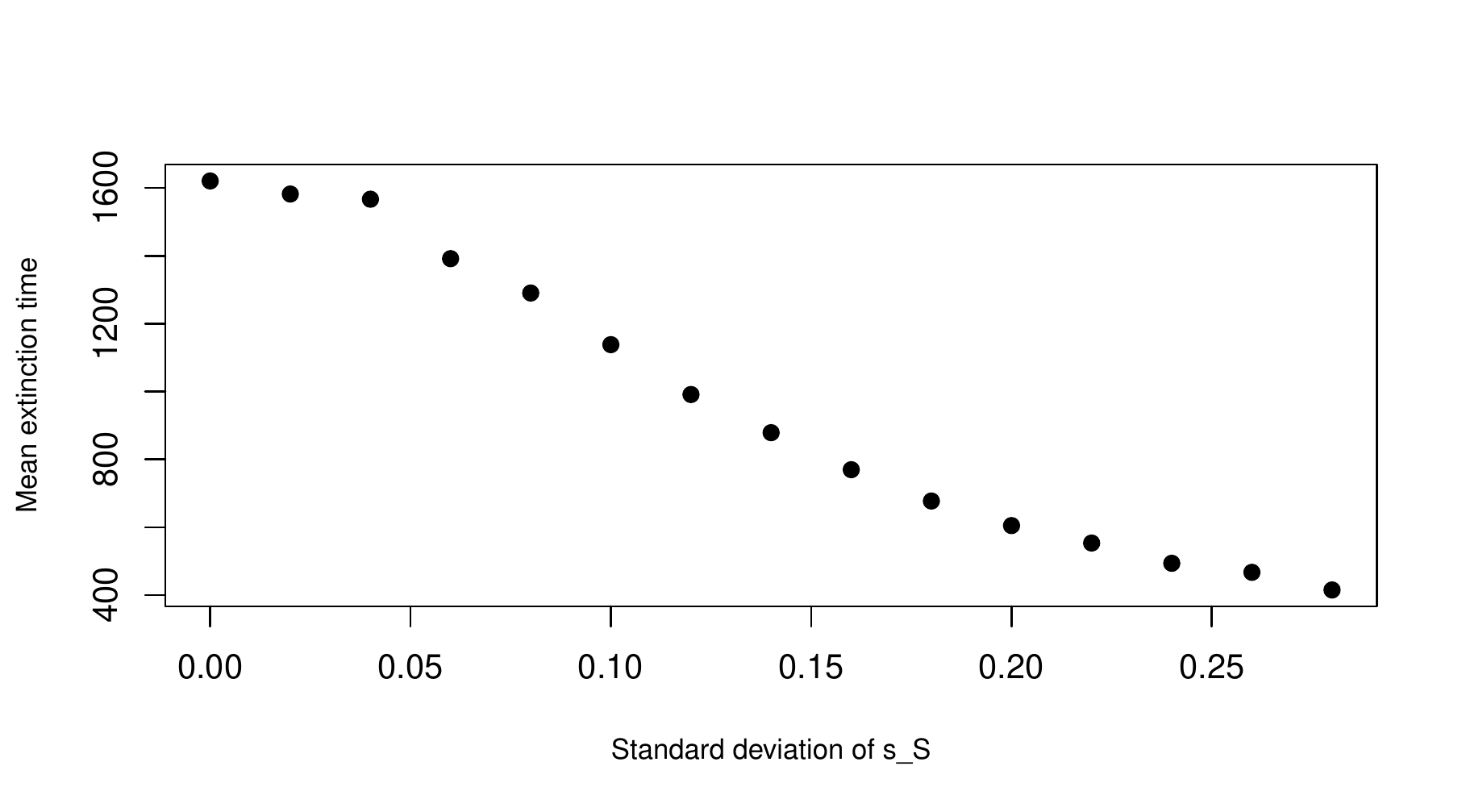}
  \caption{The mean extinction time of the branching process as a function of the standard deviation of $s_S$.}
  \label{Fig:s_S_random}
\end{figure}

We try a similar approach for another parameter. Now, the parameter $s_S$, that is, the fraction of single owls surviving to the next year, is varied. The parameter $s_S$ is now assumed to be $0.71-y$ or $0.71+y$, each with probability $1/2$, for $y=0,0.2,\ldots , 0.28$. The long-term mean of $s_S$ is still 0.71, but the variance increases. In each case, 5000 populations are simulated. The mean extinction time is plotted as a function of the standard deviation of $s_S$ in figure \ref{Fig:s_S_random}. In this case as well, we observe a monotonic decrease of the mean extinction time. 

\subsection{Variation of both parameters}
We have now introduced stochasticity in one parameter at a time, and kept the others constant. Naturally, we could try to introduce randomness in both $f$ and $s_S$. When $f = 0.56$ or $0.76$, each with probility $1/2$, the simulated mean extinction time was 1461.24. When $s_S = 0.63$ or $0.79$, each with probility $1/2$, the simulated mean extinction time was 1290.48. When we use both these random variables, independently of each other, the simulation gives a mean extinction time of 1206.67, which indicates that variation in several parameters at once may reduce the mean extinction time even more.

We cannot say whether these examples of introduced randomness are relevant with respect to the real owl population. However, the simulations show that letting the parameters be random variables instead of constants can reduce the extinction time.



\begin{thebibliography}{2}
\bibitem{Tho} J.W. Thomas, E.D. Forsman, J.B. Lint, E.C. Meslow, B.R. Noon, J. Verner, A conservation strategy for the Northern Spotted Owl, 1990-791-171/20026, U.S. Government Printing Office, Washington D.C., 1990.
\bibitem{All} L.J.S. Allen, J.F. Fagan, G. H\"ogn\"as, H. Fagerholm, Population extinction in discrete-time stochastic population models with an Allee effect, J. Difference Equ. Appl., 11 (2005), 4-5, 273-293.
\bibitem{Jun} B. Jung, Exit times for multivariate autoregressive processes, Stoch. Proc. Appl. 123 (2013), 8, 3052-3063.
\bibitem{Cou} S.P. Courtney, J.A. Blakesley, R.E. Bigley, M.L. Cody, J.P. Dumbacher, R.C. Fleischer, A.B.
Franklin, J.F. Franklin, R.J.
Gutiérrez, J.M. Marzluff, L. Sztukowski, Scientific evaluation of the status of the
Northern Spotted Owl, Sustainable Ecosystems Institute, Portland, 2004.

\end{thebibliography}
\end{document}